\documentclass[aps,prl,twocolumn,superscriptaddress]{revtex4-1}
\usepackage{graphicx}
\usepackage{bm}

\begin{document}


\title{Reply to Comment on ``Spin Ice: Magnetic Excitations without Monopole Signatures using Muon Spin Rotation"}

  \author{S.~R.~Dunsiger}
  \affiliation{Physik Department, Technische Universit\"{a}t M\"{u}nchen, D-85748 Garching, Germany}
    \author{A. A. Aczel}
     \affiliation{Department of Physics and Astronomy, McMaster University, Hamilton, Ontario L8S 4M1, Canada}
     \author{C.~Arguello}
     \affiliation{Department of Physics, Columbia University, New York, New York 10027, U.S.A.}    
  \author{H. Dabkowska}
     \affiliation{Brockhouse Institute for Materials Research, McMaster University, Hamilton, Ontario L8S 4M1, Canada}
 \author{A.~Dabkowski}
     \affiliation{Brockhouse Institute for Materials Research, McMaster University, Hamilton, Ontario L8S 4M1, Canada}
 \author{M.-H.~Du}
 \affiliation{Materials Science and Technology Division, Oak Ridge National Laboratory, Oak Ridge, TN 37831-6114, U.S.A.}
     \author{T. Goko}
     \affiliation{Department of Physics, Columbia University, New York, New York 10027, U.S.A.}
     \affiliation{TRIUMF, 4004 Wesbrook Mall, Vancouver, B.C., V6T 2A3, Canada}
     \author{B.~Javanparast}
     \author{T.~Lin}
 \affiliation{Department of Physics and Astronomy, University of Waterloo, Waterloo, Ontario, N2L 3G1, Canada}
 \author{F.~L.~Ning}
      \affiliation{Department of Physics, Columbia University, New York, New York 10027, U.S.A.}
       \affiliation{Department of Physics, Zhejiang University, Hangzhou 310027, China}
   \author{H.~M.~L.~Noad}
     \affiliation{Department of Physics and Astronomy, McMaster University, Hamilton, Ontario L8S 4M1, Canada}
 \author{D.~J.~Singh}
 \affiliation{Materials Science and Technology Division, Oak Ridge National Laboratory, Oak Ridge, TN 37831-6114, U.S.A.}
      \author{T. J. Williams}
     \affiliation{Department of Physics and Astronomy, McMaster University, Hamilton, Ontario L8S 4M1, Canada}  
    \author{Y. J. Uemura}
    \affiliation{Department of Physics, Columbia University, New York, New York 10027, U.S.A.} 
\author{M.~J.~P.~Gingras}
 \affiliation{Department of Physics and Astronomy, University of Waterloo, Waterloo, Ontario, N2L 3G1, Canada}
     \affiliation{Canadian Institute of Advanced Research, Toronto, Ontario M5G 1Z8, Canada}  
         \author{G. M. Luke}
     \affiliation{Department of Physics and Astronomy, McMaster University, Hamilton, Ontario L8S 4M1, Canada}
     \affiliation{Canadian Institute of Advanced Research, Toronto, Ontario M5G 1Z8, Canada}    
      \date{\today}
  
\maketitle

We expand on our recent publication~\cite{dunsiger11} describing the magnetic fluctuations in 
the spin ice Dy$_2$Ti$_2$O$_7$ using muon spin relaxation ($\mu$SR), prompted by work of Bramwell {\it et al.} 
published in Nature~\cite{Bramwell:2009p2963}.
Furthermore, we respond to the recent comment by Bramwell {\textit et al.}~\cite{bramwell_arx} on our publication~\cite{dunsiger11}.
The latter group claims they ``analysed a second, minority component" with a 
small characteristic internal field which showed evidence of magnetic monopoles~\cite{Ryzhkin:2005wn,Castelnovo:2008p5463}.  
Figure 1(a) of Ref.~\cite{dunsiger11} indicates $\it no$ long lived signal from the Dy$_2$Ti$_2$O$_7$ exists at 100 mK, at odds with the claim 
that ``our evidence suggests the Wien effect signal originates 
from interior muons...".  
The authors' claim the analysis of data from muons exterior to the sample was anticipated in their original proposal.  This contradicts 
Ref.~\cite{Bramwell:2009p2963}, where it is never mentioned.  Indeed, the figure captions all refer to muons {\it within} Dy$_2$Ti$_2$O$_7$.

As shown in Fig. 1, our own recent measurements at the TRIUMF facility of a 
Dy$_2$Ti$_2$O$_7$ sample mounted on silver 
qualitatively reproduce the temperature dependence of the characteristic linewidths or damping reported 
by Bramwell {\it et al.} (Fig. 4 of Ref.~\cite{Bramwell:2009p2963}).  
The data are consistently reproduced, both by ourselves as well as other authors 
studying different spin ice compounds, as pointed out by Bramwell {\it et al.}.  
The data are not in contention, but rather their interpretation.  It arises not from the magnetic 
Wien effect, but is instead a measure of the stray field in the silver surrounding the sample and 
is proportional to the bulk magnetization.  Hence we fully expect it will be evident in a 
variety of experiments on similar compounds and absent in non-spin ice samples.  If the samples 
are rotated such that muons are implanted directly into the silver sample plate, the large 
amplitude, very slowly damped signal from the latter will dominate.  

The authors of Ref.~\cite{Bramwell:2009p2963} concede themeselves the signal at T$>0.4$ K measures
the sample magnetization.  If a new minority phase developed at lower temperatures, as they claim,
it would appear as a change in signal amplitude.  As shown in Fig. 1, no such variation
is observed.  Our measurements of bare intrinsic GaAs allow us to quantitatively account
for all contributions to the $\mu$SR signal within our apparatus, from the Dy$_2$Ti$_2$O$_7$, as well as from muons exterior to the sample which constitute the background.  They rule out the existence of a minority phase.
The amplitude of the background mentioned in the Methods section of Ref.~\cite{Bramwell:2009p2963} is not quoted.  
 
The temperature variation in Fig. 1 proves the sample is in good thermal contact with the 
sample holder, contrary to their criticisms.  
As pointed out by Bramwell {\it et al.}, our spectra do not extend so far in time 
compared to their own.  The continuous nature of the 
muon beam at TRIUMF is more suited to higher timing resolution measurements on shorter timescales.
Nonetheless, a long lived signal, by definition, varies little between 5-20 $\mu$s.  Data on 
timescales $t>$6 $\mu$s are unnecessary to measure their amplitude.  

It is important to note the dimensional analysis of Bramwell {\it et al} does not give the
prefactor, or the spectral weight of the spin relaxation associated with the Wien effect.  Our own studies in a low longitudinal field configuration clearly 
show it must be dominated by other contributions.
However, irrespective of the details of dimensional analysis and the described variation of 
magnetic relaxation rate with monopole density~\cite{Bramwell:2009p2963}, in general the linewidth measured 
with $\mu$SR in a transverse field geometry
is equivalent to the free induction decay familiar from NMR.  Specifically, it is the sum of
contributions from spin-spin interactions and inhomogeneous line broadening, as well as 
spin-lattice interactions (T$_1$)~\cite{slichter}.  
These are prohibitively difficult to separate.  In a zero field (ZF)
or longitudinal field (LF) geometry, only the  T$_1$ processes due to spin fluctuations and 
dissipative energy exchange with the ``lattice" contribute.  Hence, if the 
variation in linewidth reported by Bramwell {\it et al.} arose through spin fluctuations exterior to the 
Dy$_2$Ti$_2$O$_7$, it would
be observed in a LF configuration.  As shown in Fig. 3 of Ref.~\cite{dunsiger11}, this is not the case.
A different, temperature independent process dominates, as we reported in our original 
publication~\cite{dunsiger11}.  In fact, in Ref.~\cite{giblin}, the authors specifically use ``exterior muons" and state
``the muons do not observe local magnetic fluctuations..." in such a configuration.

As pointed out by 
Blundell~\cite{blundell}, the analysis of  Ref.~\cite{Bramwell:2009p2963} only yields the theoretically 
predicted magnetic charge Q over
a very limited temperature range.  We would ask how the time dependence reported in Ref.~\cite{giblin11} may be 
differentiated from similar phenomena common to glassy systems.  However, in any case, we do not claim
magnetic monopoles do not exist within the spin ices, but rather, the experiment of Bramwell was not 
sensitive to them and further, any monopole-induced effects were swamped by other magnetic excitations.  In our 
original publication we fully acknowledged that similar temperature independent muon spin 
relaxation has been observed in a variety of geometrically frustrated magnets, as yet
poorly understood and worthy of more in depth attention.  Furthermore, we note that anomalously rapid spin fluctuations have been reported in Ising spin systems 
using a variety of techniques and are as yet incompletely understood~\cite{wolf00}.
\begin{figure}[t]
\includegraphics[angle=0,width=\columnwidth]{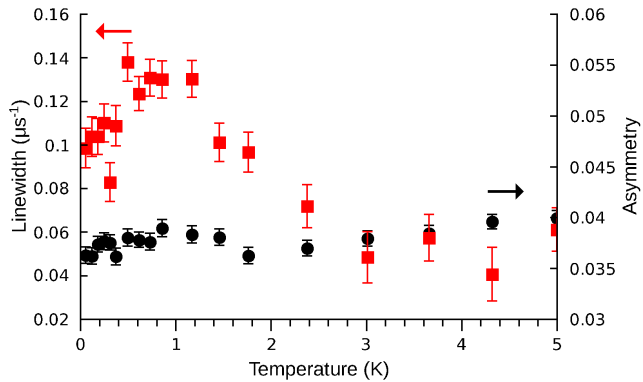} 
\caption{Linewidth and amplitude due to $\mu$SR spectra in TF= 2 milliTesla measured 
in Dy$_2$Ti$_2$O$_7$.  The contribution due to muons which miss the Ag backing plate (landing in the cryostat) has been accounted for with an additional signal.}
\end{figure}
%


\end{document}